\documentclass[pra,twocolumn,preprintnumbers,showpacs]{revtex4}
\usepackage{amsmath,amssymb,bm,graphicx}

\renewcommand\d{\partial}
\newcommand\grad{\bm{\nabla}}
\newcommand\+{\dagger}
\newcommand\<{\langle}
\renewcommand\>{\rangle}
\newcommand\x{{\bm{x}}}
\newcommand\p{{\bm{p}}}
\newcommand\q{{\bm{q}}}
\newcommand\kF{k_{\mathrm{F}}}
\newcommand\kFA{k_{\mathrm{F}\!A}}
\newcommand\kFB{k_{\mathrm{F}\!B}}
\newcommand\eFA{\varepsilon_{\mathrm{F}\!A}}
\newcommand\eff{\mathrm{eff}}
\newcommand\T{\mathcal{T}}
\newcommand{\sect}[1]{\section{#1}}

\begin{document}
\preprint{MIT-CTP 4048}

\title{Phases of a bilayer Fermi gas}
\author{Yusuke~Nishida}
\affiliation{Center for Theoretical Physics,
Massachusetts Institute of Technology, Cambridge, Massachusetts 02139, USA}

\begin{abstract}
 We investigate a two-species Fermi gas in which one species is confined
 in two parallel layers and interacts with the other species in the
 three-dimensional space by a tunable short-range interaction.  Based on
 the controlled weak coupling analysis and the exact three-body
 calculation, we show that the system has a rich phase diagram in the
 plane of the effective scattering length and the layer separation.
 Resulting phases include an interlayer $s$-wave pairing, an intralayer
 $p$-wave pairing, a dimer Bose-Einstein condensation, and a Fermi gas
 of stable Efimov-like trimers.  Our system provides a widely applicable
 scheme to induce long-range interlayer correlations in ultracold atoms.
\end{abstract}

\date{June 2009}

%03.75.Ss Degenerate Fermi gases
%05.30.Fk Fermion systems and electron gas
%67.85.Lm Degenerate Fermi gases
%74.20.Rp Pairing symmetries (other than s-wave) 
%74.78.-w Superconducting films and low-dimensional structures

\maketitle

\sect{Introduction}
One of the central themes in ultracold atoms is to provide highly
tunable model systems for other subfields in physics.  A prominent
example is the realization of ultracold two-dimensional atomic gases
which have enabled detailed microscopic studies of
Berezinskii-Kosterlitz-Thouless physics relevant to a wide variety of
two-dimensional
phenomena~\cite{Hadzibabic:2006,Schweikhard:2007,Kruger:2007}.  In
addition to simple two-dimensional systems with single layers, what have
attracted considerable attention in condensed matter physics are bilayer
or multilayer systems.  Here, extra degrees of freedom generated by
layers and the long-range Coulomb interaction between them are expected
to lead to intriguing physics such as an interlayer exciton condensation
in bilayer semiconductors~\cite{Lozovik:1975,Shevchenko:1976}, quantum
Hall bilayers~\cite{Eisenstein:2004}, and bilayer
graphenes~\cite{Kharitonov:2009}.

In ultracold atoms, the analogous multilayer geometry can be created by
confining atoms by a strong optical lattice in one direction.  However,
since the long-range Coulomb interaction is absent in neutral atoms,
separated layers are simply decoupled without interlayer tunneling.  A
novel scheme that we propose in this article to realize the long-range
interlayer correlation in ultracold atoms is to use a mixture of two
atomic species $A$ and $B$ and confine only $A$ atoms in the multilayer
geometry with keeping $B$ atoms in the three-dimensional space.
%~\cite{dipolar}.
Here the correlation between $A$ atoms confined in different layers can
be induced through the interaction with the background $B$ atoms that
are free to propagate from layer to layer. The advantage of this scheme
is that one can tune the $A$-$B$ interaction by interspecies Feshbach
resonances and thus rich phenomena are expected to occur as a function
of the interaction strength.  Furthermore, such a system may be thought
of as a tunable model system for a slab phase of nuclear matter by
identifying $A$ ($B$) atoms with protons
(neutrons)~\cite{Ravenhall:1983,Hashimoto:1984}.

\begin{figure}[bp]
 \includegraphics[width=0.95\columnwidth,clip]{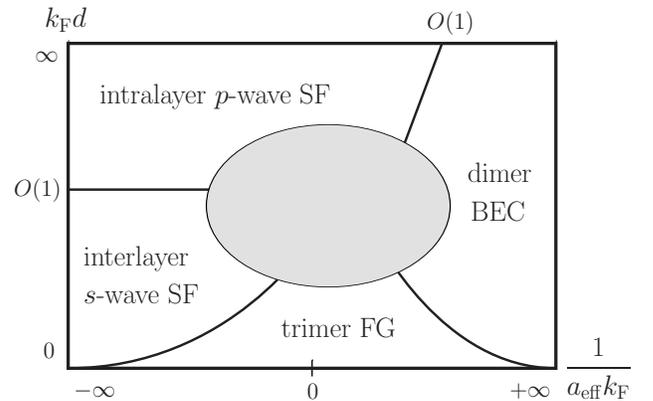}
 \caption{Conjectured phase diagram at zero temperature in the plane of
 the effective scattering length $(a_\eff\kF)^{-1}$ and the layer
 separation $\kF d$.  Here, $\kF\sim\kFA\sim\kFB$ is assumed.
 \label{fig:phase_diagram}}
\end{figure}

In this article, we give detailed analyses in the case of a bilayer
Fermi gas that can be realized by using a Fermi-Fermi mixture of, for
example, ${}^6$Li and
${}^{40}$K~\cite{Taglieber:2008,Wille:2008,Voigt:2008} and confining
only one species in two parallel layers.  Based on the controlled weak
coupling analysis and the exact three-body calculation, we show that the
system exhibits at least four distinct quantum phases as summarized in
Fig.~\ref{fig:phase_diagram}.  In the weak coupling region, an effective
attraction between two $A$ atoms mediated by $B$ atoms leads to
superfluidity due to an interlayer $s$-wave pairing (lower left phase)
or an intralayer $p$-wave pairing (upper left phase) depending on the
layer separation.  On the other hand, in the strong coupling region, an
$A$ atom captures a $B$ atom to form a tightly bound molecule and the
ground state becomes a dimer Bose-Einstein condensation (right phase).
Finally, when the $A$-$B$ interaction is close to the resonance, we find
that two $A$ atoms confined in different layers with one $B$ atom form a
three-body bound state leading to a Fermi gas of trimers in a dilute
system (lower middle phase).  Remarkably, there is an infinite number of
such three-body bound levels exactly at the resonance resembling the
Efimov effect in three dimensions~\cite{Efimov:1970}.

%\newpage

The system under consideration is described by the action (here and
below $\hbar=1$ and $k_\mathrm{B}=1$):
\begin{equation}\label{eq:action}
 \begin{split}
  & S = \sum_{i=1,2}\int\!dtd\x\,\psi_{Ai}^\+(t,\x)
  \left(i\d_t+\frac{\grad_{\!\x}^2}{2m_A}+\mu_A\right)\psi_{Ai}(t,\x) \\
  & + \int\!dtd\x dz\,\psi_B^\+(t,\x,z)
  \left(i\d_t+\frac{\grad_{\!\x}^2+\nabla_{\!z}^2}{2m_B}+\mu_B\right)\psi_B(t,\x,z) \\
  & + g_0\sum_{i=1,2}\int\!dtd\x\,
  \psi_{Ai}^\+(t,\x)\psi_B^\+(t,\x,z_i)\psi_B(t,\x,z_i)\psi_{Ai}(t,\x).
 \end{split}
\end{equation}
Here, $\psi_{Ai}(t,\x)$ with $\x=(x,y)$ is a spinless fermionic field for
$A$ atoms confined in two parallel layers that are labeled by the index
$i=1,2$ and located at $z=z_i$.  $\psi_B(t,\x,z)$ is another spinless
fermionic field for $B$ atoms in the three-dimensional space.  The last
term in the action describes the short-range $A$-$B$ interaction.  $g_0$
is a cutoff ($\Lambda$) dependent bare coupling, which can be eliminated
by introducing the physical parameter, the effective scattering length
$a_\eff$, through
$\frac1{g_0}-\frac{\sqrt{m_Bm_{AB}}}{2\pi}\Lambda=-\frac{\sqrt{m_Bm_{AB}}}{2\pi{}a_\eff}$
with $m_{AB}\equiv m_Am_B/(m_A+m_B)$ being the reduced
mass~\cite{Nishida:2008kr,Nishida:2008gk}.  $a_\eff$ is arbitrarily
tunable by means of the interspecies Feshbach resonance and the limit
$a_\eff\to-(+)0$ corresponds to the weak (strong) attraction between $A$
and $B$ atoms.  Throughout this article, the interlayer tunneling is
assumed to be negligible and Fermi momenta of $A$ and $B$ atoms are
defined through their densities; $\kFA\equiv(4\pi n_A)^{1/2}$ and
$\kFB\equiv(6\pi^2n_B)^{1/3}$.

\begin{figure}[tp]
 \includegraphics[width=0.95\columnwidth,clip]{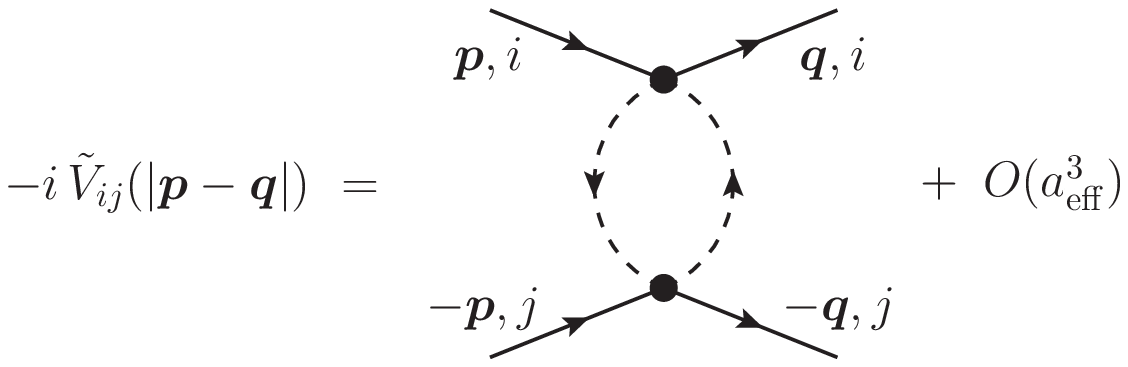}
 \caption{Effective interaction between two $A$ atoms (solid line)
 induced by the interaction with $B$ atoms (dotted line) in the weak
 coupling limit.  Here, $i,j=1,2$ are layer indices.
 \label{fig:induced}}
\end{figure}

\sect{Weak coupling limit}
In order to elucidate phases appearing in our system, we start with the
weak coupling limit $a_\eff\to-0$ in which a controlled perturbative
analysis is possible.  Unlike the ordinary BCS-BEC crossover purely in
two or three dimensions, the $A$-$B$ pairing at weak coupling does not
take place in our mixed dimensional system because of the absence of the
Cooper instability~\cite{Nishida:2008gk}.  What can happen instead are
pairings between $A$ atoms using the effective attraction induced by the
interaction with the Fermi sea of $B$ atoms.  To the leading order in
$a_\eff$, the back-to-back scattering of $A$ atoms in layers $i$ and $j$
is described by the Feynman diagram in Fig.~\ref{fig:induced}.  The
resulting induced interaction $\tilde V_{ij}(|\p-\q|)$ has rather a
simple form in the real space:%~\cite{Nishida:2008ra}:
\begin{equation}\label{eq:induced}
 V_{ij}(|\x|) = \frac{a_\eff^2}{m_{AB}}
  \frac{2\kFB r_{ij}\cos(2\kFB r_{ij})-\sin(2\kFB r_{ij})}{4\pi r_{ij}^4},
\end{equation}
where $r_{ij}=\sqrt{\x^2+(z_i-z_j)^2}$ is the distance between the two
$A$ atoms.  The oscillatory decaying factor is well known in the
Ruderman-Kittel-Kasuya-Yosida
interaction~\cite{Ruderman:1954,Kasuya:1956,Yosida:1957} and leads to
the following physical interpretation of $V_{ij}(|\x|)$:  The density
modulation of background $B$ atoms produced by one $A$ atom in the layer
$i$ mediates the long-range interaction with the other $A$ atom in the
layer $j$.

Now the pairings of $A$ atoms are described by the BCS-type Hamiltonian:
\begin{align}
 & H_\mathrm{ind} = \sum_i\sum_\p\tilde\psi_{Ai}^\+(\p)
 \left(\frac{\p^2}{2m_A}-\mu_A\right)\tilde\psi_{Ai}(\p) \\
 & + \frac1{2\Omega}\sum_{i,j}\sum_{\p,\q} \notag
 \tilde\psi_{Ai}^\+(\p)\tilde\psi_{Aj}^\+(-\p)
 \tilde V_{ij}(|\p-\q|)\tilde\psi_{Aj}(-\q)\tilde\psi_{Ai}(\q).
\end{align}
In the mean-field approximation, the gap function is given by
$\Delta_{ij}(\p)=\frac1\Omega\sum_\q\tilde V_{ij}(|\p-\q|)\<\tilde\psi_{Aj}(-\q)\tilde\psi_{Ai}(\q)\>$
and the Fermi statistics of $A$ atoms implies
$\Delta_{ji}(-\p)=-\Delta_{ij}(\p)$.  Although $A$ atoms are spinless
fermions, the layer indices $i,j=1,2$ make more than one pairing pattern
possible.  When the layer separation is large
$d\equiv|z_1-z_2|\to\infty$, the interlayer interaction $V_{i\neq j}$ is
suppressed while the intralayer interaction $V_{i=j}$ is unaffected by
$d$.  Therefore, in this limit, intralayer pairings with
$\Delta_{11},\Delta_{22}\neq0$ are favored while their pairing symmetry
has to be $p$~wave~\cite{Nishida:2008gk}.  On the other hand, when the
layer separation becomes small $d\to0$, $V_{i\neq j}$ is no longer
suppressed and thus an interlayer pairing with
$\Delta_{12}=-\Delta_{21}\neq0$ is favored because the pairing symmetry
can be $s$~wave only in this singlet channel.  Therefore, there has to
be a quantum phase transition as a function of the layer separation.

The quantum phase transition can be located by comparing the energy
densities for the interlayer $s$-wave and intralayer $p$-wave pairings.
In the weak coupling limit, their energy densities are, respectively,
given by
$\<\mathcal{H}_\mathrm{ind}\>=-\frac{m_A\mu_A^2}{2\pi}-\frac{m_A|\Delta_{12}|^2}{4\pi}$
and
$\<\mathcal{H}_\mathrm{ind}\>=-\frac{m_A\mu_A^2}{2\pi}-\frac{m_A|\Delta_{11}|^2}{8\pi}-\frac{m_A|\Delta_{22}|^2}{8\pi}$,
where the gap functions at $|\p|=\kFA$ are given by
$\Delta_{12}/\eFA\propto e^{2\pi/[m_A\tilde V_{12}^{(0)}]}$ and
$\Delta_{11}/\eFA=\Delta_{22}/\eFA\propto\left(\hat{p}_x\pm i\hat{p}_y\right)e^{2\pi/[m_A\tilde V_{11}^{(1)}]}$
for negative $\tilde V_{ij}^{(l)}$.  Here, $\eFA\equiv\kFA^2/(2m_A)$ is
the Fermi energy and 
$\tilde V_{ij}^{(l)}\equiv\int_0^\pi\!\frac{d\varphi}{\pi}\cos(l\varphi)\tilde V_{ij}(|\p-\q|)$
with $\cos\varphi\equiv\hat{\p}\cdot\hat{\q}$ is the partial wave
projection of the induced interaction in which both incoming and
outgoing momenta are restricted on the Fermi surface of $A$ atoms;
$|\p|=|\q|=\kFA$.  It is the phase with the smaller energy density and
hence the larger pairing gap that is realized in our system.

\begin{figure}[tp]
 \includegraphics[width=0.85\columnwidth,clip]{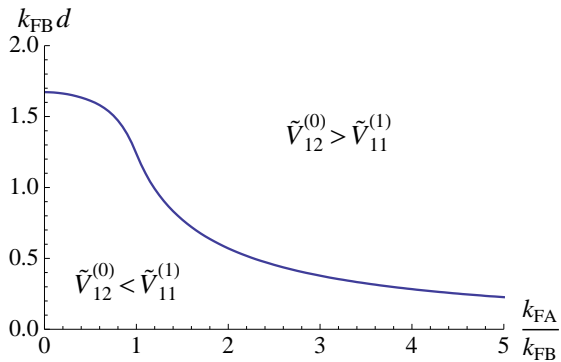}
 \caption{(Color online) Critical layer separation $\kFB d$ dividing the
 interlayer $s$-wave pairing
 $\bigl(\tilde V_{12}^{(0)}<\tilde V_{11}^{(1)}\bigr)$ from the
 intralayer $p$-wave pairing $\bigl(\tilde V_{12}^{(0)}>\tilde
 V_{11}^{(1)}\bigr)$ as a function of $\kFA/\kFB$.  \label{fig:pairing}}
\end{figure}

The critical layer separation dividing the above two phases is plotted
in Fig.~\ref{fig:pairing} as a function of $\kFA/\kFB$, which is
independent of $a_\eff$ and $m_A/m_B$ to the leading order in $a_\eff$.
Below the critical separation where
$\tilde V_{12}^{(0)}<\tilde V_{11}^{(1)}$, the interlayer $s$-wave
pairing appears, while the intralayer $p$-wave pairing appears above the
critical separation where $\tilde V_{12}^{(0)}>\tilde V_{11}^{(1)}$.
The two layers exhibit superfluidity in both phases and its critical
temperature at weak coupling is the same order as the pairing gap at
zero temperature; $T_\mathrm{c}\sim|\Delta_{ij}|$.  We note that our
interlayer $s$-wave pairing bears an analogy to the interlayer exciton
condensation in condensed matter
systems~\cite{Lozovik:1975,Shevchenko:1976,Eisenstein:2004,Kharitonov:2009}.
Also it is worthwhile to point out that the intralayer $p_x{+}ip_y$-wave
pairing has a potential application to topological quantum computation
using vortices with non-Abelian
statistics~\cite{Read:2000,Tewari:2007}.

\sect{Strong coupling limit}
We now turn to the strong coupling limit $a_\eff\to+0$, in which $A$
atoms confined in layers capture $B$ atoms from the bulk space to form
two-body bound states (dimers) whose binding energy is given by
$E_2=-\frac1{2m_{AB}a_\eff^2}$.  The resulting system consists of the
dimers localized around layers which interact weakly with each other and
excess $B$ atoms.  As long as the dimer size $\sim a_\eff$ is smaller
than the layer separation $d$ and the mean interparticle distance
$\sim\kFA^{-1}$, the dimers behave as two-dimensional bosons and
therefore the ground state becomes a dimer Bose-Einstein condensation in
each layer.  Accordingly, the two layers exhibit superfluidity up to the
Berezinskii-Kosterlitz-Thouless temperature given by
$T_\mathrm{BKT}\to\frac{2\pi n_d}{M}\ln^{-1}\!\left(-\frac{380}{4\pi}\ln{n_da_\eff^2}\right)$
in the limit $a_\eff\to+0$~\cite{Prokefev:2001}, where $M\equiv m_A+m_B$
and $n_d=n_A$ are the dimer's mass and density per layer.  We note that
the dimer Bose-Einstein condensates in different layers can interact
with each other through the excess $B$ atoms just as in
Fig.~\ref{fig:induced} and Eq.~(\ref{eq:induced}).

Although our system is found to exhibit superfluidity in both weak and
strong coupling limits $a_\eff\to\mp0$, their symmetry breaking patterns
are different.  The action in Eq.~(\ref{eq:action}) has continuous
symmetries of
$\mathrm{U}(1)_{A1}\times\mathrm{U}(1)_{A2}\times\mathrm{U}(1)_B\times\mathrm{U}(1)_z$
corresponding to particle number conservations of $A$ atoms in each
layer, that of $B$ atoms, and a rotation about $z$ axis.  An order
parameter of the interlayer $s$-wave pairing
$\<\epsilon_{ij}\psi_{Ai}\psi_{Aj}\>\neq0$ breaks the full symmetries
down to
$\mathrm{U}(1)_{A1-A2}\times\mathrm{U}(1)_B\times\mathrm{U}(1)_z$, while
order parameters of the intralayer $p_x{+}ip_y$-wave pairing
$\<\psi_{Ai}(\tensor\d_x+i\tensor\d_y)\psi_{Ai}\>\neq0$ break them down
to $\mathrm{U}(1)_{A1+A2-z}\times\mathrm{U}(1)_B$.  Order parameters of
the dimer Bose-Einstein condensation are
$\<\psi_{A1}\psi_B\>,\<\psi_{A2}\psi_B\>\neq0$, which break the full
symmetries down to $\mathrm{U}(1)_{A1+A2-B}\times\mathrm{U}(1)_z$.
Therefore, the phases in the weak and strong coupling limits have to be
divided by at least one quantum phase transition as a function of the
interaction strength.  Below we show that a novel phase can appear in
between when the layer separation is small.

\sect{Unitarity limit}

\begin{figure}[tp]
 \includegraphics[width=\columnwidth,clip]{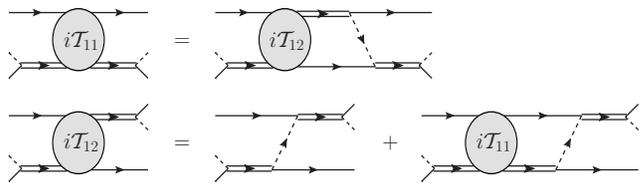}
 \caption{Three-body scattering of two $A$ atoms in different layers and
 one $B$ atom.  The double line represents the $A$-$B$ scattering
 amplitude.  $\T_{22}$ and $\T_{21}$ satisfy the same integral equations
 as $\T_{11}$ and $\T_{12}$.  \label{fig:3-body}}
\end{figure}

The fourth phase realized in our system can be elucidated by studying a
three-body problem of two $A$ atoms confined in different layers 
interacting with one $B$ atom.  Their scattering process is depicted in
Fig.~\ref{fig:3-body} and all the relevant diagrams can be summed by
solving the integral equations for $\T_{ij}(E;\p,\q)$.  Here $E$ is the
total energy in the center-of-mass frame and $\p$ ($\q$) is the relative
momentum of the incoming (outgoing) $A$ atom in the layer $i$ ($j$) that
does not scatter with the $B$ atom first (last).  The $\T$-matrix
elements have properties $\T_{11}=\T_{22}$ and $\T_{12}=\T_{21}$ and can
be decomposed into even- and odd-parity parts under the exchange of one
layer index; $\T_\pm\equiv\T_{11}\pm\T_{12}$.  After the partial wave
projection
$\T_\pm^{(l)}(E;p,q)\equiv\int_0^\pi\!\frac{d\varphi}{\pi}\cos(l\varphi)\T_\pm(E;\p,\q)$
with $\cos\varphi\equiv\hat{\p}\cdot\hat{\q}$, we find $\T_\pm^{(l)}$ to
satisfy
\begin{equation}\label{eq:integral}
 \begin{split}
  & \T_\pm^{(l)}(E;p,q) = \mp \sqrt{m_Bm_{AB}}\,K^{(l)}(E+i0^+;p,q) \\
  & \mp \int_0^\infty\!dk\,k\frac{\T_\pm^{(l)}(E;p,k)\,K^{(l)}(E+i0^+;k,q)}
  {\sqrt{\frac{m_B+m_{AB}}{M}k^2-2m_{AB}E-i0^+}-\frac1{a_\eff}},
 \end{split}
\end{equation}
where $K^{(l)}(E;p,q)$ is given by
\begin{equation}
 \int_0^\pi\!\frac{d\varphi}{\pi}\cos(l\varphi)
  \frac{e^{-d\sqrt{\frac{M}{m_A}}
  \sqrt{p^2+q^2+\frac{2m_A}{M}pq\cos\varphi-2m_{AB}E}}}
  {\sqrt{p^2+q^2+\frac{2m_A}{M}pq\cos\varphi-2m_{AB}E}}.
\end{equation}

\begin{figure}[tp]
 \includegraphics[width=0.85\columnwidth,clip]{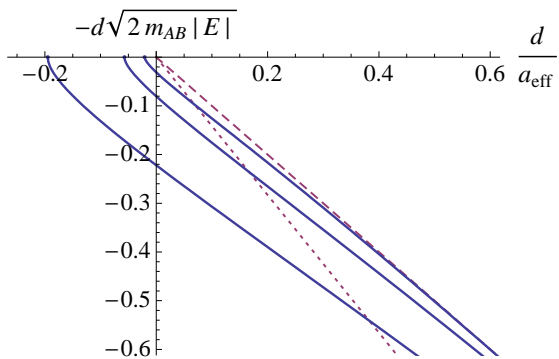}
 \caption{(Color online) Lowest binding energies of trimers for mass
 ratios $m_A/m_B=6.67$ (bottom), $1$ (middle), and $0.15$ (top) as a
 function of $d/a_\eff$.  The dashed and dotted lines are atom-dimer and
 dimer-dimer thresholds; $E=E_2$ and $2E_2$.  \label{fig:binding}}
\end{figure}

The spectrum of three-body bound states (trimers) is obtained by poles
of $\T_\pm^{(l)}$ as a function of $E$.  When $E$ approaches one of the
binding energies $E_3<-\frac{\theta(a_\eff)}{2m_{AB}a_\eff^2}$, we can
write $\T_\pm^{(l)}$ as
$\T_\pm^{(l)}(E;p,q)\to\mathcal{Z}_\pm^{(l)}(p,q)/(E-E_3)$.  By solving
the homogeneous integral equation from Eq.~(\ref{eq:integral}) satisfied
by the residue $\mathcal{Z}_\pm^{(l)}$, we find that the trimers exist
only in an odd-parity ($-$) and $s$-wave ($l=0$) channel.  The lowest
binding energy $E_3^{(0)}$ as a function of $d/a_\eff$ is plotted in
Fig.~\ref{fig:binding} for three different values of the mass ratio
$m_A/m_B=0.15$, $1$, and $6.67$, corresponding to a mixture of
$A={}^6$Li and $B={}^{40}$K, two different internal states of the same
atomic species, and $A={}^{40}$K and $B={}^6$Li.  The trimer appears on
the negative side of the effective scattering length $|a_\eff|\sim d$
and its binding energy decreases as $d/a_\eff$ is increased.  When
$d/a_\eff\to+\infty$, the trimer binding energy asymptotically
approaches the atom-dimer threshold $E=E_2$ from below as
$E_3/E_2-1\propto e^{-\frac{m_B+m_{AB}}{M}\exp\left(\sqrt{\frac{M}{m_A}}\frac{d}{a_\eff}\right)}\to+0$.

Interestingly, when the $A$-$B$ interaction is exactly at the two-body
resonance $|a_\eff|\to\infty$, there exists an infinite number of such
trimer states whose spectrum is expressed by the form
$E_3^{(n)}=-e^{-2\pi n/s_0}\frac{\kappa^2}{2m_{AB}d^2}$ for $n\to\infty$
(not shown in Fig.~\ref{fig:binding}).  Such a geometric spectrum at the
resonance is well known as the Efimov effect in three
dimensions~\cite{Efimov:1970}.  Here the scaling exponent $s_0$ and the
so-called Efimov parameter $\kappa$ can be determined as
$(s_0,\kappa)=(0.741,0.0311)$, %(0.741141,0.0310713)
$(0.828,0.0807)$, %(0.827847,0.0807289)
$(1.30,0.231)$ %(1.29775,0.230801)
for $m_A/m_B=0.15$, $1$, $6.67$, respectively.  An important characteristic
of our trimer state in contrast to Efimov trimers in a free space is its
stability against the three-body recombination because the two $A$ atoms
are spatially separated.

The existence of stable $AAB$ trimers, which has the same quantum number
as $\epsilon_{ij}\psi_{Ai}\psi_{Aj}\psi_B$, can lead to a Fermi gas of
trimers in a dilute system.  As long as the trimer size
$(2m_{AB}|E_3|)^{-1/2}\sim d$ is smaller than the mean interparticle
distance $\sim\kFA^{-1}$, the trimers behave as two-dimensional fermions
localized around layers and form a Fermi gas with its density equal to
$n_A$.  Although the trimer state exists even in the strong coupling
limit $a_\eff\to+0$, the trimer gas phase cannot persist there because
an addition of another $B$ atom breaks up the trimer into two dimers
when $E_3>2E_2$ is reached (see the dimer-dimer threshold in
Fig.~\ref{fig:binding}).  Therefore, the trimer Fermi gas is realized
only when the $A$-$B$ interaction is close to the resonance
$|a_\eff|\gtrsim d$.
We note that $AAB$ trimers with $A$ atoms in the same layer and $ABB$
trimers are absent for $0.0351<m_A/m_B<6.35$~\cite{Nishida:2008kr}.
Once they are formed, the whole phase diagram would be dominated by such
deeply bound trimers whose size is set by the thickness of layers.

\sect{Conclusions}
We found that the dimer Bose-Einstein condensation appears in the strong
coupling region where $a_\eff\lesssim d$ and $a_\eff\kFA\lesssim O(1)$,
while the trimer Fermi gas appears in the unitarity region where
$|a_\eff|\gtrsim d$ and $\kFA d\lesssim O(1)$.  Assuming that the rest
of the phase diagram is occupied by the interlayer $s$-wave or
intralayer $p$-wave superfluidity found in the weak coupling region, we
conclude that the system has the rich phase diagram shown in
Fig.~\ref{fig:phase_diagram}.  Our results can be tested by using a
ultracold Fermi-Fermi mixture of ${}^6$Li and ${}^{40}$K with a
species-selective optical lattice.  How the above four phases meet at
the center of the phase diagram (shaded region in
Fig.~\ref{fig:phase_diagram}) and whether more quantum phases appear in
our system are interesting open questions and can be addressed, in
principle, by future experiments.  In particular, further study on how
the dilute gas of trimers evolves into a gas of their constituents as
$\kFA d$ is increased may shed light on a similar problem of the
transition from nuclear matter to quark matter existing in the core of
neutron stars.  Finally, we emphasize that the scheme to induce
long-range interlayer correlations in ultracold atoms that is presented
in this article by taking the bilayer Fermi gas as an example can be
widely extended to multilayer geometries, multiwire geometries,
Bose-Bose mixtures, and Bose-Fermi mixtures.

\acknowledgments
Y.\,N.\ is supported by MIT Pappalardo Fellowship in Physics and
the US Department of Energy Office of Nuclear Physics under Grant
No.\ DE-FG02-94ER40818.

\end{document}